\documentstyle[12pt,eqsecnum,aps,epsf]{revtex}
\newcommand{\etal}{\mbox{\it et al.}}
\tighten

\begin{document}

\preprint{~}

\title{Radio-Frequency Measurements of \\
Coherent Transition and Cherenkov Radiation:\\
Implications for High-Energy Neutrino Detection}

\author{
Peter~W.~Gorham$^{3}$,
David~P.~Saltzberg$^{2}$, 
Paul~Schoessow$^{1}$, 
Wei~Gai$^{1}$, 
John~G.~Power$^{1}$, 
Richard~Konecny$^{1}$, 
and M.~E.~Conde$^{1}$}
\address{$^{1}$Argonne National Laboratory}
\address{$^{2}$Department of Physics and Astronomy,
University of California, Los Angeles}
\address{$^{3}$Jet Propulsion Laboratory, Calif. Institute of Technology,
Pasadena, CA, 91109}

\date{\today}

\maketitle
\begin{center}
({\it To be published in Physical Review E})
\end{center}
\begin{abstract}
We report on measurements of 11-18~cm wavelength radio emission
from interactions of 15.2~MeV pulsed electron bunches 
at the Argonne Wakefield
Accelerator. The electrons were observed both 
in a configuration where they produced primarily transition radiation
from an aluminum foil, and in a configuration designed for the electrons
to produce
Cherenkov radiation in a silica sand target.   Our aim was to emulate 
the large electron excess expected to 
develop during an electromagnetic cascade initiated by an
ultra-high-energy particle.  Such charge asymmetries are predicted
to produce strong coherent radio pulses, which are the basis for
several
experiments to detect high-energy neutrinos from the showers they induce 
in Antarctic ice and in the lunar regolith. 
We detected coherent emission which we attribute both to transition
and possibly
Cherenkov radiation at different levels depending
on the experimental conditions.  We discuss implications for
experiments relying on radio emission for detection of electromagnetic
cascades produced by ultra high-energy neutrinos.

\end{abstract}
\pacs{41.75.Fr,98.70.Sa,41.60.Bq}
%41.75.Fr  Electron & Positron beams
%98.70.Sa  Cosmic Rays
%41.60.Bq  Cherenkov Radiation

%>>>>>>>>>>>>>>>>>>>>>>>>>>>>>>>>>>>>>>>>>>>>>>>>>>>>>>>>>>>>>>>>>>
%>>>>>>>>>>>>>>>>>>>>>>>>>>>>>>>>>>>>>>>>>>>>>>>>>>>>>>>>>>>>>>>>>>
\section{Introduction}

In 1962, G. Askaryan\cite{Ask62} first predicted that large 
electromagnetic particle cascades
would develop an asymmetry of electrons over positrons at 
the level of 20-30\% due to the following processes:
Compton scattering of atomic
electrons in the target material by cascade photons, 
$\delta$-ray scattering processes, 
and annihilation of positrons
in flight. Askaryan realized that the propagation of this
charge asymmetry in a dielectric medium should
result in radio emission due to the processes of
Cherenkov and transition radiation.   For wavelengths much larger
than the shower dimensions, he noted that the radio emission 
would be coherent, {\it i.e.},
the electrons radiate in phase, with a resulting 
$N_e^2$ increase in the radiated
power with the number of excess electrons $N_e$, in contrast with
a linear rise expected from incoherent emission.

Recently this suggestion has generated renewed interest both on 
experimental and theoretical fronts. 
On the experimental side, it has become the basis for a number of new
searches for radio-frequency (RF) emission from cascades 
induced by ultra-high energy neutrinos---at PeV energies in Antarctic ice
\cite{bol91,fri96,Seck99,Bes99}
and at EeV energies from the lunar regolith
\cite{Alv96,Han96,Zhe88,Dag89,Gor99}. 
On the theoretical side, a number of recent, very detailed 
simulations\cite{Alv96,Zas92,Alv97,Alv98}
have validated Askaryan's basic result:
Particle cascades at energies from $\sim$1~TeV up to $10^{20}$~eV
will develop a net negative charge excess of order
20-30\% of their total charged-particle number. These
simulations also confirm Askaryan's prediction of
a corresponding nanosecond pulse of radio Cherenkov
emission, with a characteristic frequency and angular dependence.
Extending Askaryan's original idea, Markov and Zheleznykh~\cite{Mark86}
have also shown that a number of other coherent radiation processes,
such as transition, synchrotron, and bremsstrahlung radiation, are
likely to be important contributors to the total emission from
cascades depending on the cascade medium, ambient fields, and geometry.

\subsection{Experimental tests}

Under typical conditions, none of these processes produce radio emission
comparable to that of other passbands until the
energy of the cascades exceeds $\sim 10^{15}$~eV. Thus a direct
accelerator beam test of these predictions is difficult.   Such a test
could be made with existing TeV proton beams where observations of
radio emission, though
expected to be correspondingly 
weak and difficult to characterize, would demonstrate the existence of the
charge asymmetry. However, a failure to detect radio emission under
these circumstances might be attributable to either a failure of the shower
to produce the predicted charge asymmetry, or a loss of radio coherence
in the emission process after the charge asymmetry has developed.

As an alternative experimental approach, we have utilized lower energy
pulsed electron bunches incident on a target material. The electron 
bunch then emulates the resulting charge
asymmetry from the late stages of a high energy cascade, and allows
one to focus experimentally on the radio emission process. Simulations
have shown\cite{Zas92} that the bulk of the radio emission arises from
low-energy electrons ($\leq 10$ MeV) produced late in the shower. Thus
investigations of the coherent radio emission properties can be done 
with fairly modest electron energies, provided that the bunch length
is at least as small as the 
longitudinal and transverse shower sizes expected in cascades in solid
materials (several~cm).  The total number of charged particles,
(which are virtually all electrons and positrons near and after shower 
maximum)
in a cascade is to a good approximation $E_{shower}/\mbox{GeV}$.
Hence, a given cascade energy can be simply related to the number of 
electrons per pulse at an accelerator.

We have adopted this alternative approach to begin developing a
laboratory basis for confirmation of Askaryan's hypothesis. Using
the Argonne Wakefield Accelerator (AWA) facility at Argonne National
Laboratory, we have employed $20-50$ ps pulsed electron bunches with 
$N_e \simeq 10^{10-11}$ electrons per bunch to produce coherent radio
emission from several targets.  We looked for radio emission
in the 1.7--2.6 GHz microwave
band (11-18 cm wavelength).  Our initial measurements show evidence 
for both coherent transition and perhaps
Cherenkov radiation. In the following
section we discuss some of the history of related measurements at 
other facilities.  Section~\ref{theory} gives a summary of the theoretical 
basis for the emission.  In section~\ref{setup} we describe our experimental
setup, the measurements made, and the results obtained.  We 
conclude the paper by considering the applications
of these results to coherent radio detection of neutrino-induced cascades.

\subsection{Related measurements of coherent emission from LINAC beams}

Over the last two decades a significant effort has gone into
measurements of coherent Cherenkov radiation (CR)
and transition radiation (TR)
from electron linear accelerator (LINAC) beams.  
The primary motivation for these experiments has been the
development and characterization of coherent sources of millimeter-wave and 
far-infrared radiation.
Although several early 
experiments~\cite{Nei84,Maru86} were done at centimeter wavelengths using
continuous-beam devices, most recent work has focussed on the shorter
wavelengths.  
In the early cm-wave work
the detected power was attributed to
CR from air at the end of the beam pipe.
No measurements of the coherence properties were made.
 
Later work~\cite{Nak89,Shi91,Ish91}
demonstrated the coherence of radiation observed under similar
conditions, again attributing it in part to CR.
Following this, a number of authors\cite{Shi91,Tak94,Shi94}
realized that the bulk
of the observed radiation in such circumstances must in fact be
primarily TR, and that the Cherenkov contribution
was not easily separable given the experimental configurations used.
This is primarily due to two features of these experiments:
First, the Cherenkov angular distribution in air is nearly identical to that 
of TR.  Second, the relatively short effective path-lengths of the 
electron beam in the air meant that the CR power was suppressed 
compared to the TR power.

\section{Theoretical considerations}
\label{theory}

As we have noted above,
the primary radio emission processes that we expect to be 
important in our experiment are transition and Cherenkov radiation.
Both of these processes are well-known and studied in the optical
(for CR) and X-ray (for TR) regimes, but the emission is not typically
coherent at these wavelengths. In addition, most CR emission is 
observed under conditions where the emitting region and the
charged-particle track length is physically much longer than the
wavelength observed; for our accelerator experiments
this condition is not satisfied and
the traditional treatment of CR through the Frank-Tamm formulation,
which obtains when the track length is much longer than the wavelength
of observation,
is not valid.  In the case of transition radiation, most present
work centers on the use of X-ray TR as a diagnostic for particle
energies in detector systems at accelerators or in cosmic-ray
detectors. A number of the approximations used for X-ray TR are
not valid for radio emission, and our treatment here reflects that.

\subsection{Transition radiation}

Transition radiation occurs as relativistic
electrons cross the boundaries between dielectric media.
The forward angular spectrum of TR for passage of a single
electron through a single dielectric
boundary with dielectric constants and indices of refraction
$\epsilon_1,n_1$ upstream and
$\epsilon_2,n_2$ downstream is given by~\cite{Ginz90,Tak94}:
\begin{equation}
\label{tr-eq}
\frac{d^2 W_{TR}}
     {d\omega d\Omega} ~=~
\frac{\hbar \alpha}
     {\pi ^{2}}
\frac{\sqrt{\epsilon_2}\sin^2\theta\cos^2\theta}
     {(1-\beta^2 \epsilon_2 \cos^2\theta)^2}
\ |\zeta|^{2}
\end{equation}
where the angle $\theta$ is measured with respect to the electron
direction, $\beta=v/c$ is the electron velocity, and $\hbar$ and $\alpha$
are Planck's constant (over $2\pi$) and the fine-structure constant,
respectively. The factor $\zeta$ is
\begin{equation}
\zeta ~=~
\frac{(\epsilon_2 - \epsilon_1)(1-\beta^2\epsilon_2-\beta\sqrt{\epsilon_1-
      \epsilon_2\sin^2\theta})}
     {(\epsilon_1\cos\theta +\sqrt{\epsilon_2}\sqrt{\epsilon_1-\epsilon_2\sin^2\theta})
      (1-\beta\sqrt{\epsilon_1-\epsilon_2\sin^2\theta})}
\end{equation}
The equation is written this way since the factor $|\zeta \cos\theta|$ is
close to unity for most solid--vacuum interfaces.
We have neglected the magnetic permeability of the two media
since it is unimportant for the materials in our experiment.
The quantity expressed is the total radiated energy per unit 
radian frequency per unit solid angle. The dielectric constants $\epsilon_i$
are in general complex with both refractive and absorptive components.
These equation show that
the TR is forward-peaked with
a characteristic angle $1/\gamma$ and that the radiation 
is broadband. 
%with equal number of quanta per unit bandwidth.
Analogous
results can be obtained for the backwards emission 
by multiplying by the appropriate Fresnel reflection coefficient. 
We refer the
reader to the literature~\cite{Ginz90,Garib,War75} for details.

We note that
these results are derived for the case of a particle traversing
from $-\infty$ and do not account for deceleration of the particle
at the interfaces or in the media.  These equations
contain poles at several values of $\theta$ and $\epsilon_2$,
most notably at the value of $\theta$ defined by
\begin{equation}
\cos \theta =  (\beta \sqrt{\epsilon_2})^{-1},
\end{equation} 
which is the 
condition normally associated with the definition of the Cherenkov
angle. This has been identified as an artifact of the assumption of an
infinite medium and track length~\cite{Ginz90,Garib},
but it does highlight the
close association of the TR and CR processes, and 
the complications that arise in separating them.
As a result, equation~\ref{tr-eq} does not provide
a clear theoretical distinction which would allow us to separate
out the assumptions of an infinite medium and track length, in
order to isolate the TR from the CR contribution. We
discuss this issue further in appendix~\ref{examples}.

Another important feature of transition radiation is its 
{\em formation zone} $L_f$ ~\cite{Ginz90}, which is the downstream region of
the medium over which the radiation field due to the
transition becomes fully separated from the coulomb field of the
propagating charge. For fully-formed TR
to develop, the dielectric boundary must be sharp compared
to the size of this region, given by
\begin{equation}
L_f ~=~ { 2\pi\beta c \over |\omega (1- n_2\beta\cos\theta )|}.
\end{equation}
This also means that any effect which disturbs the dielectric properties
of the downstream medium within the formation zone will tend to
suppress the full formation of TR according to the equations developed 
above. 
For the case where
the particle is stopped within the formation zone, the assumption
used in developing the TR formalism are again not satisfied and
we expect that the TR is again suppressed. 
At present the magnitude of this expected suppression
is not quantified in the literature, but we note this effect since
it is likely to be important at some level in our experiment.

\subsection{Cherenkov radiation}

Cherenkov radiation from a finite electron track length is treated
by Tamm\cite{Tam39}, and discussed more recently in the context
of coherent radiation from a LINAC beam by Takahashi $\etal$\cite{Tak94}
For electrons traversing a track of length $L$ in a medium
of index of refraction $n$~\cite{Tak94}:
\begin{equation}
\label{cr-eq}
{d^2W_{\mbox{CR}} \over d\lambda d\Omega} ~=~ { h c \alpha n \over \lambda^2}
\left ( { L \over \lambda }\right )^2 \mbox{sinc}^2X(\lambda,\theta)~ 
\sin^2\theta .
\end{equation}
Here $\mbox{sinc} \ x \equiv \mbox{sin} \pi x / \pi x$ and
\begin{equation}
X(\lambda,\theta) ~=~ { L \over \beta\lambda} (1-\beta n \cos\theta),
\end{equation}
where $\lambda$ is the wavelength of the radiation. 
The quantity expressed is the total radiated energy per unit 
wavelength per unit solid angle.   As a result, the CR is peaked
at a characteristic {\em Cherenkov angle} 
$\cos \theta_C =(\beta n)^{-1}$. 
%and,
%with equal radiated energy per unit bandwidth, is even more broadband than
%TR.

As we have noted above, the treatments of TR in the literature often
contain elements that appear as limiting cases of a Cherenkov-like
contribution. In fact similar statements can be made about treatments
of Cherenkov radiation for the case of a finite track length.
Takahashi $\etal$\cite{Tak94} have shown that this relation can be 
interpreted in terms of a CR-like component from the
continuous portion of the track added to TR-like components from the
end-points of the track.  We have not attempted to make such
a distinction, but we note again the strong physical connection
between TR and CR.

\subsection{Coherent emission from electron bunches}

The equations presented above have not explicitly treated the
effects of coherence on the emission process. 
There is however an extensive literature describing the coherence
properties of radiation from electron bunches\cite{Var75,Shi94,Tak94}.
As long as the transverse size of the bunch is less than a fraction
of the wavelengths observed, the electric fields of each radiating
electron add in phase and the net electric field grows linearly
with the number of electrons, implying a quadratic growth of power.
Thus for $N_e$ electrons per bunch, we expect the total radiated
energy be $N_e^2$ times the formulas given above for single electrons.

This is the condition for full coherence, and it will only obtain in the
limit where the bunch size is far smaller than the wavelength observed.
In our case, although this condition is generally satisfied, there
are corrections due to the bunch form factor which should be 
accounted for. This is typically done by introducing the form factors
for the longitudinal, transverse, and angular emittance of the
electron bunch:
\begin{equation}
P ~=~ N_e(1+ N_e f_L f_T f_{\chi}) P_0~.
\end{equation}
Here $P$ is the radiated power, $f_L,f_T,f_{\chi}$ are the 
longitudinal, transverse, and angular emittance form factors, and
$P_0$ the single-particle radiated power\cite{Tak94}. The spatial form
factors can be estimated via a three-dimensional Fourier transform
of the electron distribution (here assumed to have cylindrical
symmetry); the angular form factor is somewhat more complex but
can also be numerically evaluated if an estimate of the 
angular emittance of the beam is available\cite{Tak94}.

\section{Experiment}
\label{setup}

Our experiment, although limited in part by some of the same
effects as earlier works in separating coherent TR
from CR, has been designed so that the Cherenkov angular distribution
is quite distinct from the
forward-peaked TR distribution. To do this we
use silica sand as our target medium. This has the further advantage
of being a more relevant medium with respect to the ongoing
high energy neutrino experiments, since the material is comparable in
refractive index and RF attenuation properties to both ice and
the lunar regolith. It has the disadvantage of stopping a
lower-energy electron
shower relatively quickly, and this tends to weaken the resulting
CR emission relative to that of TR, because of the $L^2$
dependence of the CR power on path length evident in equation
\ref{cr-eq} above.

\subsection{Electron beam}

The Argonne Wakefield Accelerator~(AWA)\cite{Conde} 
at Argonne National Laboratory provided our electron bunches.
AWA produces  electron bunches of mean energy 15.2 MeV, and a bunch length
of $\sim 1$ cm. The transverse distribution at the
exit to the beam pipe was Gaussian with 
$\sigma \approx 7$ mm.
The beam was pulsed once per second with a beam
current tunable from 0.5 to 25~nC/pulse.
The beam current was measured bunch-by-bunch
using an integrating current transformer
(ICT).  The ICT produced an output voltage pulse proportional
to the measured bunch charge which was recorded event-by-event.

\subsection{Target}

\begin{figure}
\begin{center}
%\centerline{\psfig{file=tgeom.eps,height=3in}}
%\vspace{10pt}
\leavevmode
\epsfxsize=5in
\epsfbox{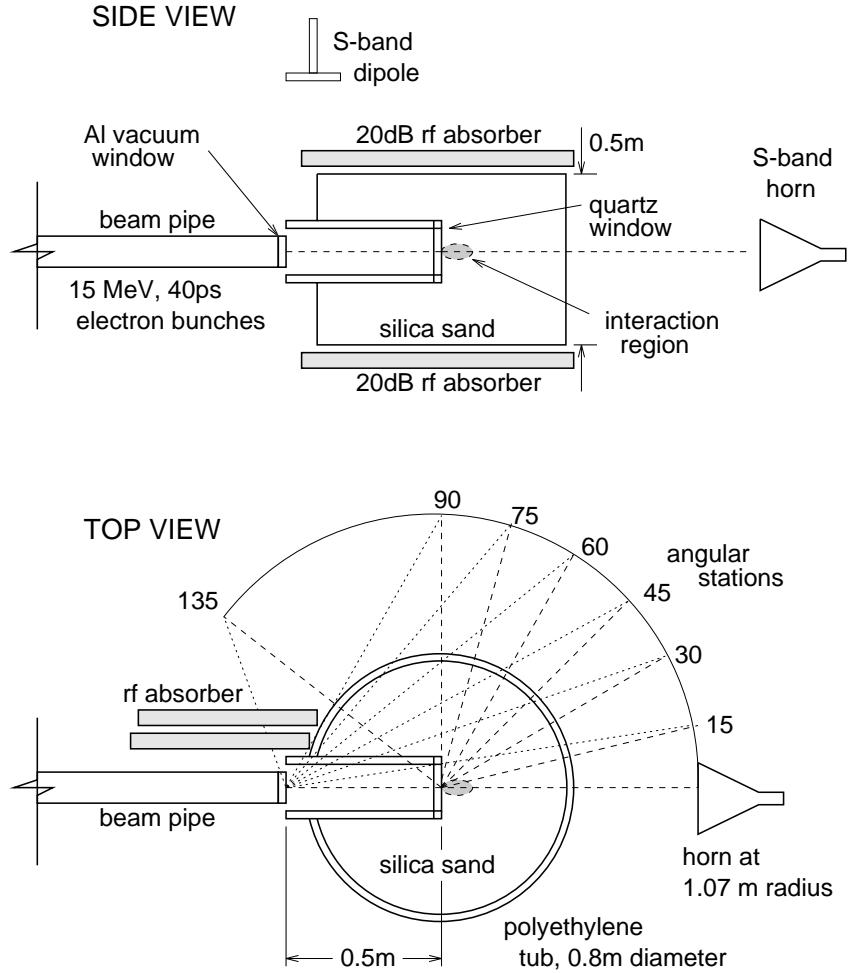}
\caption{Two sectional views of the target geometry.}
\label{tgeom}
\end{center}
\end{figure}

The target geometry is shown in Fig.~\ref{tgeom}.
The target consisted of a 80~cm diameter by 55~cm high
polyethylene tub (wall thickness $\sim$3~mm)
with a 6 inch diameter, schedule 40 PVC pipe
penetrating its side. 
We used an empty tub for our initial runs with the PVC pipe left open
on both ends.
Thus the
beam exited the accelerator through a 3 mil aluminum window and ranged out
in air several meters beyond the experiment.   

In subsequent runs, we filled the tub
with 360 kg of silica sand, to within 7 cm of its top.
The grain sizes were fairly uniform with 300~$\mu$m diameters.  The
average density (including the packing factor) was 1.58~g/cm$^{3}$.
We measured the index of refraction of the sand at 2~GHz to be
$1.55\pm 0.01$, indicating that the water content was less than 3\% by 
weight.~\cite{hippel} The attenuation of microwaves by 1~m of
such dry sand is small, amounting to a loss of about 1.3 dB at 2 GHz.
For the  sand runs, the PVC pipe was capped with a quartz
window at the internal end to hold in the sand. 
Our original intent was to lead the vacuum beam pipe into the center of
the target but this proved impractical due to difficulties in 
adequately supporting the beam pipe. Thus the
the electron bunches were
transmitted through the aluminum vacuum window at the end of
the beam pipe, through about 50 cm air, and entered the
target through a quartz window of about 6 mm thickness.

We placed 20~dB of
RF absorber above and below the tub to minimize reflections.
To suppress
RF emission from the TR at the beampipe into the backward direction,
we generally ran with 20 dB of RF absorber along the side of
the beampipe as shown in Fig.\ref{tgeom}.

\subsection{Simulation of expected electron shower}

\begin{figure}
\begin{center}
\vspace{0.3cm}
\leavevmode
\epsfxsize=4.5in
\epsfbox{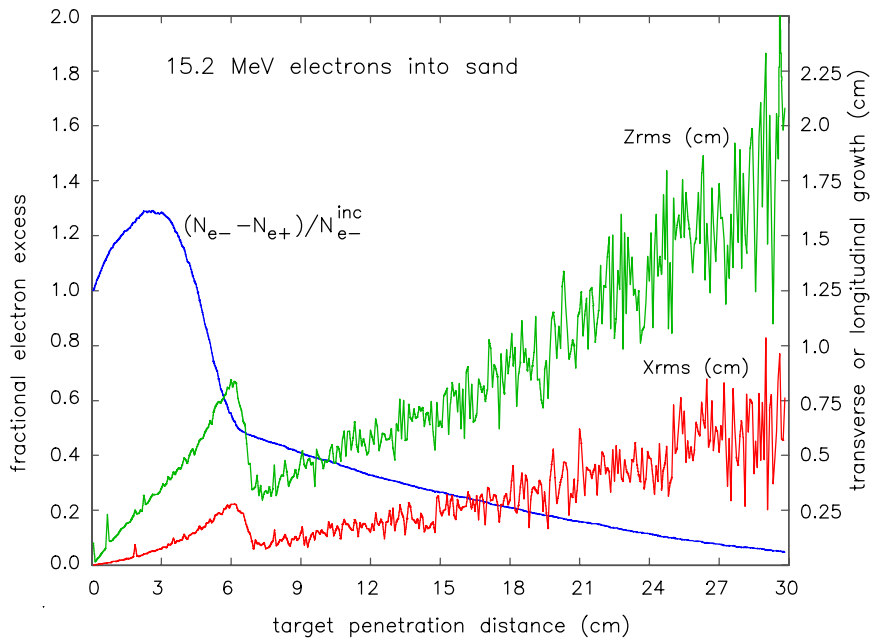}
\vspace{0.3cm}
\caption{Simulation of the expected electron shower
distribution for the 15.2 MeV electron bunch from the AWA
facility, here done for quartz 
scaled for the lower density of sand used. The fractional
number curve represents the net charge, accounting for
the positrons produced. 
The change in rms bunch
dimensions during propagation through the target is
denoted by Zrms (longitudinal) and Xrms (transverse).}
\label{sio2a}
\end{center}
\end{figure}

To characterize the expected shower behavior
it was simulated using the EGS4 Monte Carlo~\cite{slac}
assuming that the material was quartz, but with the
density adjusted to match that of the sand.
This allowed us to estimate the longitudinal development
in electron number as well as the elongation and lateral growth of the shower.
The results of one such simulation are shown in Fig.~\ref{sio2a}. 
Here the curve marked $(N_{e-}-N_{e+})/N_{e-}^{inc}$ shows 
the evolution of the 
electron number as a small shower develops in the initial few cm,
and then dies out, with a long tail of stragglers. The other
two curves show the expected growth in the rms size in both transverse
(Xrms) and longitudinal (Zrms) directions.
Based on these results, we expect the CR emission in the sand to 
be primarily from the $\sim 3-4$ cm region of the shower maximum, with
some additional contribution due to Compton
electrons extending out to $\sim 20$ cm or more. This latter
contribution is reduced due to the increase in the bunch size which
decreases the coherence of radiation from this portion of the
electron shower.

\subsection{Antennas}

\begin{figure} 
\begin{center}
%\centerline{\psfig{file=hornbeam.eps,height=3in}}
%\vspace{10pt}
\leavevmode
\epsfxsize=4.in
\epsfbox{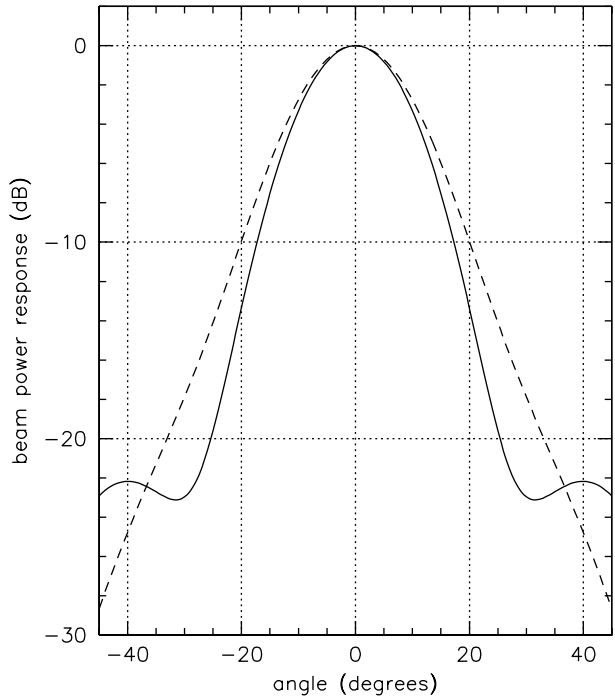}
\caption{Calculated angular response lobes of the pyramidal horn.  The
solid and dashed lines are for the H- and E-plane, respectively.}
\label{lobes}
\end{center}
\end{figure}

We received the RF emission signal using a
pyramidal horn, with a nominal half-power bandpass of 
1.7--2.6~GHz and directivity of 15.3 dB at
2.15 GHz.  The rectangular aperture of the horn was 36.5~cm (E-plane) by
27.3~cm (H-plane).  The angular response of the antenna is shown in 
Fig.~\ref{lobes}.
Phase errors at the front of the horn relative to its throat cause
the effective area of a horn to be less than its geometric area.
We calculated the ratio of effective aperture to geometric
aperture to be 0.51, which is typical for such horns.  Using a network
analyzer we estimated further inefficiencies due to impedance mismatches
and ohmic losses to be of order 10\%.  We estimate our ability to point the
horn to be 5$^{\circ}$, corresponding to a 2~dB loss on average.

We used a balanced half-wave 6.9~cm dipole with a 
half-power bandwidth of $\sim 700$ MHz
centered at 1.8 GHz to provide a trigger.
The dipole signal had a risetime of $\sim 150$~ps and 
provided a trigger stable to 40~ps or better.

The antenna positions
are shown in Fig.~\ref{tgeom}.  The dipole was
50 cm from the exit of the beampipe at an angle of $70^{\circ}$ from
the beam axis, 
and was stationary during the course
of the experiment.   The horn was moved
along a fixed radius from the target center to a series of fixed
angular positions from 0$^{\circ}$ to 135$^{\circ}$. 
The outputs of the two antennas were fed without
amplification into 20~m of coaxial cables 
which brought the signals 
into the control room with a total attenuation of 6--7~dB.

As is typical for an accelerator experiment which must be done within the
confines of a shielded vault, we were limited in our ability to place
the antennas further than about 1-2~m from the source of the radiation.
Thus the response of the horn will be 
affected by near-field (Fresnel zone) effects.
The primary effect within the Fresnel zone is a loss of antenna 
efficiency due to wavefront curvature which produces a
phase error across the receiving aperture.  
%Since a pyramidal horn
%is typically designed to excite a single TEM mode, only those portions
%of the aperture that contribute to this mode will be efficiently 
%coupled to the horn.  
Using calculations in Ref~\cite{jull}, we estimate
these effects to cause a 0.25-0.4~dB power loss, depending on the
angle of the horn.

\subsection{Data acquisition}

All measurements of pulsed RF emission were done in the time domain 
using a Tektronix 694C real-time
digital sampling oscilloscope with
3~GHz bandwidth and 10~GS/s,~8-bit linear digitization of four channels.
The signals from the horn and dipole required no amplification; in fact,
because of the strength of the coherent radio emission, induced
voltages in the antennas had to be limited by
RF attenuators (typically 20 dB), 
to bring the signal into an acceptable
range for the $50~\Omega$ scope inputs.
The scope produces a time series of the
voltages interpolated from 100~ps to 40~ps resolution, which is slightly
faster than the input channel risetime, measured by
Tektronix to be in the range of 50-60~ps. 
Thus we were able to sample the full bandwidth
of the antenna outputs and thereby make direct measurements of
the electric field intensities, mediated only by the response of
the antennas and cables used.  
The excellent
time resolution was useful for
identifying spurious reflections from structures near the target.
For each measurement, between 12 and 50 triggers were recorded. All
of the recorded pulse profiles were far above ambient RF noise
levels, and the pulse-to-pulse variation in the profiles is typically
less than 5\% per sample.

\subsection{Power and voltage measurements}

Detected pulse energy at the face of the antenna 
was calculated by summing $4V^{2}/(50~\Omega)$ ($V$ is the measured voltage)
over a 3~ns window (75 bins) defined by the primary radio pulse
from the electron bunch, either from the Al window of the
beampipe (for the TR runs) or the target center (for the CR runs).
During this time window the power
contributions from spurious reflections were negligible.
The factor of 4 accounts for voltage dividing between the antenna and load.
We measured the 3~ns time window to contain 98\% of the total pulse energy as
measured at the scope.  The origin of the window was shifted to account
for different propagation delays in the sand and air (1~ns).
The raw recorded voltages were corrected directly for the effects of
attenuators, cable and adaptor losses, near-field losses, 
transmission loss, aperture
inefficiency, and pointing errors described above.

\subsection{Datasets}

All of the data presented in this paper were taken over a two-day 
period, Sept. 23--24, 1999.  Except for the ``pure TR runs'', the
measurements were taken 107.3~cm from the center of the tub with the
horn pointing at the tub center.

\begin{itemize}

\item\underline{Pure TR runs}:
To establish the baseline TR contribution without possible CR effects
from the target, we took data with the tub removed so that the only
significant radiation would be TR from the aluminum beampipe vacuum window.
Measurements were made with the horn at 8.5 and 16.6$^{\circ}$, pointing
directly at the aluminum window.  
The horn was placed 183~cm from the foil for these runs.

\item\underline{Empty-target runs}:
Additional TR measurements were made using an empty target without
the quartz window in place. Our initial goal here was to allow
for the possibility of subtraction of the TR signal from the
target-full runs; thus a complete set of angular measurements were
made, under the same configurations as the full-target runs:
that is, with the horn always pointing toward the target center.
However, since there was neither any sand nor the quartz window
present, the observed emission received by the horn actually
emanates from the beampipe end. Emission of TR from the beampipe 
end is thus received off-axis by the horn at a given angular 
position, and must be corrected for the known beam response of the horn,
which we have done. 

Measurements were taken at 15$^{\circ}$ intervals
from 15 to 135$^{\circ}$.  In addition measurements at the 45$^{\circ}$
position were repeated at a range of beam currents to measure the coherence
properties of the TR in our band.

\item\underline{Full-target runs}:
With sand in the target, the electrons produce TR as they pass
through the beampipe end, then both TR and CR as the enter the target.
TR emission from the end of the beampipe will undergo refraction
through the target. Refraction is less of an issue for 
CR and TR emission from the
sand and quartz window, because we chose the cylindrical geometry to 
minimize such effects. However, any assessment of 
full-target contributions of the strong TR emission
due to the beampipe end must account for
for the refractive effects of the target.
  
Measurements were taken at 15$^{\circ}$ intervals
from 0 to 135$^{\circ}$. For a subset of angles we also measured
the polarization of the radiation.

\item\underline{Diagnostic runs}:
We took several special runs with strategically placed absorber sheets
to allow for comparison studies of different
portions of the observed time structure of the pulses.  For example,
we placed an absorber over the portion of the target viewed by the
horn to allow us to identify multipath reception which was
not directly 
associated with the target.  Thus we identified unwanted reflections
and eliminated them from consideration in the analysis.

\end{itemize}

\section{Results}

\subsection{TR measurements}

To provide the simplest geometry for subsequent analysis of the
TR emission, we made several runs without the large target present.
To identify any possible emission from
stray surface currents, diffraction,
or image charge effects that
might be induced by the beam along the beampipe, we made
identical measurements in this sequence both
with and without a $0.7\times0.7$~m grounded 
piece of heavy Al foil
several cm downstream from the vacuum window.
The thickness of the foil (several mil) was well above the
skin depth at our frequencies, and was intended to provide
a TR radiator plane much larger than the wavelengths
of interest.

\begin{figure}
\begin{center}
%\centerline{\psfig{file=TR0.eps,height=6.2in,width=4.6in}}
%\vspace{10pt}
\leavevmode
\epsfxsize=4.5in
%\epsfbox{TR0.eps}
\epsfbox{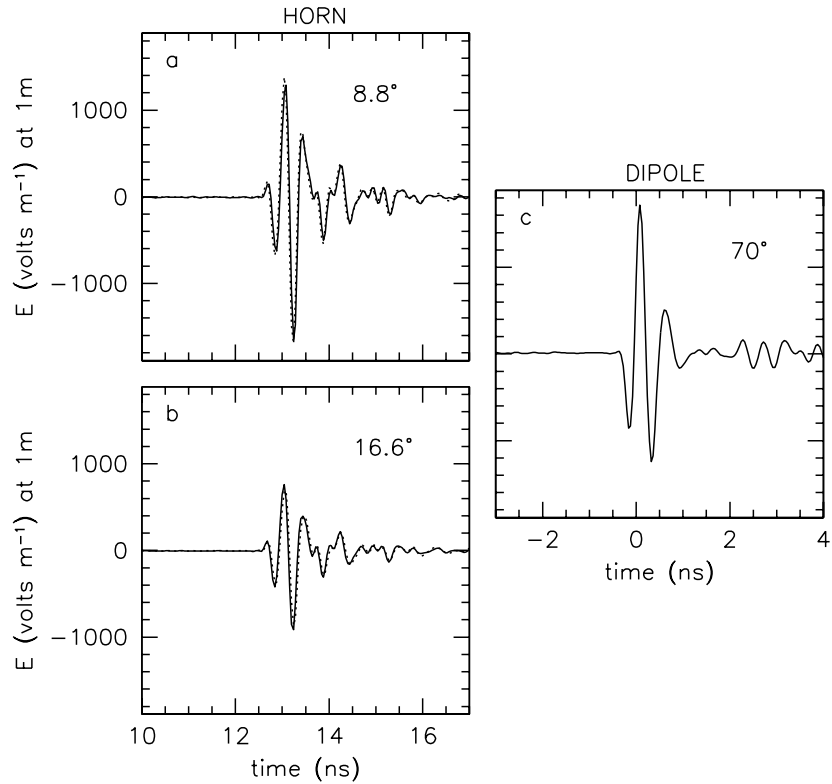}
\caption{Typical horn and dipole 
measured field strengths from TR produced at the beampipe end.
(a),(b) Field strength from the horn are shown for two angles, with
(c) showing the corresponding strength 
measured at the dipole, which was fixed in angle at $70^{\circ}$. 
The solid and dotted lines in (a) and (b) 
are from runs with and without a foil shield as described
in the text.}
\label{tr0}
\end{center}
\end{figure}

Fig.~\ref{tr0} shows the measured field strengths for the two angles.  
The field strengths are corrected for attenuation
and referenced to a distance of 1 m. Conversion of the measured
voltages to field strengths requires a knowledge of the effective height of
the antennas used; we estimated these to be 18 cm for the horn, and
7 cm for the dipole.
The time axis is relative to
the dipole trigger, shown in (c), 
which had a shorter cable length and was
closer to the source.  The emission with the large foil 
present is shown in solid lines, and that
without the foil in the dotted lines. There is essentially no
difference between the two. We conclude that there was no
contribution due to unexpected image charge or other effects from the
beampipe end.

\begin{table}
\caption{Measured and calculated pulse energies for the TR
runs.  \label{trtable}}
\vspace{0.2cm}
\begin{tabular}{c|c|c|c|c}
antenna& angle & measured energy & calculation  & meas./calc.\\
& ($^{\circ}$) & ($\mu$J/sr)  & ($\mu$J/sr) &  \\
\hline
horn & 8.5  &5.8 & 156.0 & 0.037\\
horn & 16.6 &1.7 & 45.5 & 0.037\\
dipole & 70 & 21.6 & 3.3 & 6.25 \\
\end{tabular}
\end{table}

There is a marked difference in the pulse shapes between the
dipole and horn, which is likely to be due both to differences
in the effective bandwidth of the two antennas and to partial
impedance mismatches in the cables. However, in both cases  
the full-width at half maximum (FWHM)
of the pulses is of order 1 ns, which indicates that
the detected pulses are band-limited, since $\Delta T \simeq
1~{\rm ns} \simeq (\Delta f)^{-1}$. 
This is 
consistent with the view of TR as a process which arises very rapidly
as the electrons cross a dielectric boundary, in this case from
aluminum to air. 

One might expect that there is some contribution to the measured pulse
energy from
CR along the air path within our acceptance angle.
However, because the effective pathlength for CR production is 
fairly short in this case, and the microwave index of refraction of the
air is close to unity, the CR contribution can be neglected
with respect to the TR.

The fully corrected pulse energy measurements and those predicted by 
Equation~\ref{tr-eq}, where we have assumed that the real part of
the Al dielectric constant is $\epsilon_r = 10$ at 2 GHz, 
are shown in Table~\ref{trtable}.  
For the horn measurements, the two are
in disagreement by a factor of nearly 30. 
For the dipole, our single measurement
at an angle of $70^{\circ}$ appears to give us more power than
expected from the theory.  As yet, the source of these
discrepancies is not completely understood, and resolution of this
issue is the subject of further work. We do however, expect that
the systematic uncertainty in our absolute calibration is as much as
50\%, which can account for a portion, though not all,
of the difference. 

\subsection{Coherence of TR}

\begin{figure} 
\begin{center}
%\centerline{\psfig{file=allcoh.eps,height=5.5in,width=4.6in}}
%\vspace{10pt}
\leavevmode
\epsfxsize=4.5in
\epsfbox{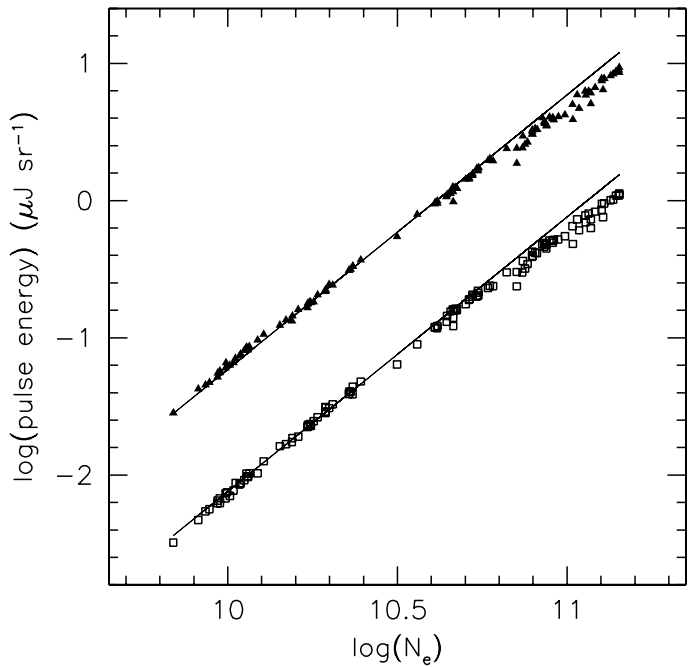}
\caption{Detected energy per pulse as a 
function of beam current. The upper
points correspond to the dipole and the lower points the 
horn. All power measurements have been normalized to 15~nC electron bunches.
In this configuration, the
primary radiation is expected to be transition radiation from the
Al vacuum window.  The solid lines are the
$N_e^2$ behavior if full coherence obtains, normalized to the
lowest beam currents. }
\label{trcoh}
\end{center}
\end{figure}

A crucial feature of this radio emission necessary for
its application to detection
of high energy particles is its coherence.
Figure~\ref{trcoh} shows the behavior of the detected power as
a function of beam current for the no-target configuration where
the radiation is expected to be primarily TR.  Here the top 
points are from the dipole measurements, again at an angle of 
70$^{\circ}$ from the beampipe end. The lower points are
for the horn measurements, here all taken at an
angle of 45$^{\circ}$ from the beampipe end.  
The solid lines have a slope of 2, 
representing the expected
behavior for power proportional to $N_e^2$ (full coherence).

The quadratic
dependence of power with the number of electrons per pulse is
evident, although there is some deviation from this at the
highest pulse charges. It is unknown at this time whether this
deviation is intrinsic (for example, from space-charge
effects) or is due to non-linearity in the ICT.

\subsection{TR angular power distribution}

\begin{figure} 
\begin{center}
%\centerline{\psfig{file=mtpwr.eps,height=4.8in,width=4.2in}}
%\vspace{10pt}
\leavevmode
\epsfxsize=4.5in
\epsfbox{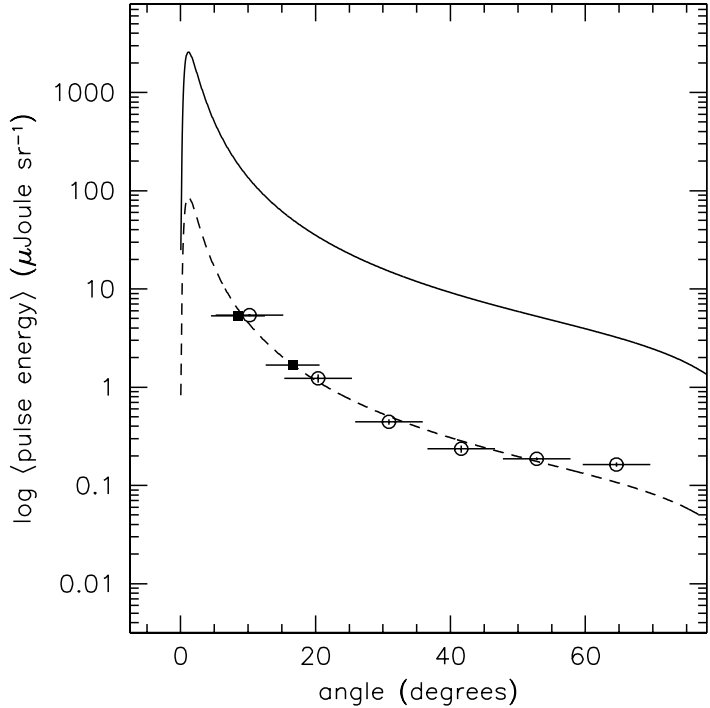}
\caption{Average microwave pulse 
energy plotted as a function of the angle for the target-empty case.
Here we have corrected the energy for the beam response and the
differing distances.  All power measurements have been normalized to
15~nC electron bunches.
The error bars are statistical in the vertical
direction and represent the half-power angular acceptance of the horn in the
horizontal direction. The solid curve is the expected distribution
for TR from the aluminum vacuum window.}
\label{mtpwr-fig}
\end{center}
\end{figure}

In Figure~\ref{mtpwr-fig} we show a plot of the angular
power spectrum measured by the horn
for runs where we expect TR to dominate: the original low-angle
runs with no target present (solid points), 
and the empty-target runs (open points).
The power for the latter case has been
corrected for the beam response of the horn (see Fig~\ref{lobes})
and for the varying distances of the antenna position with respect to the
beampipe end. The vertical error bars are statistical, and the horizontal
bars show the acceptance angle of the horn. The data are plotted
according to the effective angle with respect to the beampipe end.
As we noted previously, Cherenkov emission from the air path does not
contribute significantly in this case.
 
The plotted solid curve shows the expected theoretical
angular distribution for 
TR, assuming an Al-to-air interface, and
the dielectric constant as noted above.
The curve shows the overall discrepancy in pulse energy with
respect to the data
we have noted above in Table 1.
The dashed curve is the same TR distribution, now
normalized to the data value at $8.5^{\circ}$. The data shows
reasonable agreement with the expected angular dependence,
with some probable systematic differences between the target-absent
runs and the target-empty runs.

\subsection{Comparison of target empty to target full}

\begin{figure}
\begin{center} 
%\centerline{\psfig{file=trcrfl0.eps,height=6.2in,width=5.3in}}
%\vspace{10pt}
\leavevmode
\epsfxsize=5.5in
\epsfbox{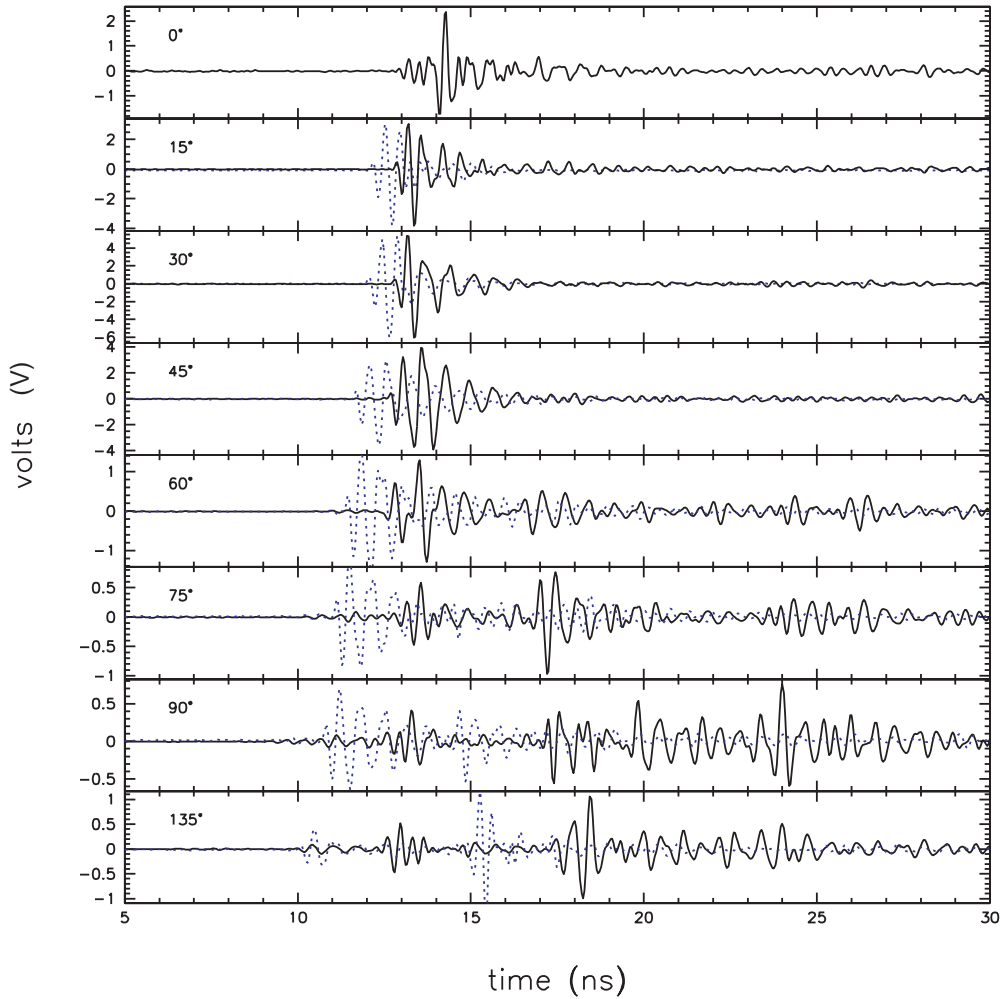}
\vspace{0.3cm}
\caption{Typical traces of horn response versus angle.  The pulses
from the sand remain in phase as the angle is changed, consistent with
power emanating from the center of the target.  The angular dependence
of the time delay for the empty target is shown for comparison in
dotted lines.}
\label{fl-fig}
\end{center}
\end{figure}

In figure~\ref{fl-fig} we plot an angular sequence of profiles
of the received pulses from the target.   
The solid lines correspond to pulse profiles with the sand in,
and the voltage scale at left corresponds to these profiles.
The dotted lines show the pulse profiles for the same
sequence of pulses for the case when
the target was present but had not yet been filled with
sand, and the voltages have been scaled arbitrarily to fit them
into the corresponding target-full profile.
No measurement at $0^{\circ}$ was
made in this case since the antenna would have been within the beam.
All pulse features later than $\sim 16$ ns in the target full 
profiles (later than $\sim 15$ ns in the overlain target-empty
profiles) are due to reflections from objects near the target,
as we confirmed with diagnostic runs during the experiment.
In each case,
the trigger is identical, and the absolute timing
is preserved to a small fraction of 1 ns. Thus all timing 
differences between the plots for different angles or
for the target-empty vs. target-full case arise from 
differences in the target geometry and the position of the
source of the radiation.

There are two obvious distinctions between the traces for full target (solid)
and empty target (dotted) in Figure~\ref{fl-fig}.
First, the ``full'' profiles at low angles appear about 0.8 ns later than
the corresponding ``empty'' profiles. This is expected since the 
refractive index of the sand induces a delay for any radiation that 
propagates through it, whether it originates at the end of the
beampipe or from the center of the target. The index of refraction
for silica sand at these frequencies is expected to be in the 
range of 1.5-1.7 depending on water content; our data indicates a 
value  1.55 as noted above.

Second, the target-full profiles in general remain centered around 
13 ns, rather than showing the progression to
earlier times that is evident in the target-empty data. This
indicates that the emission in the target-full case
arises from near the center of the target,
since the horn antenna position was at a fixed radius from this 
point. The delay of these pulses is also consistent with the
geometry.

\begin{figure}
\begin{center} 
\leavevmode
\epsfxsize=4in
\epsfbox{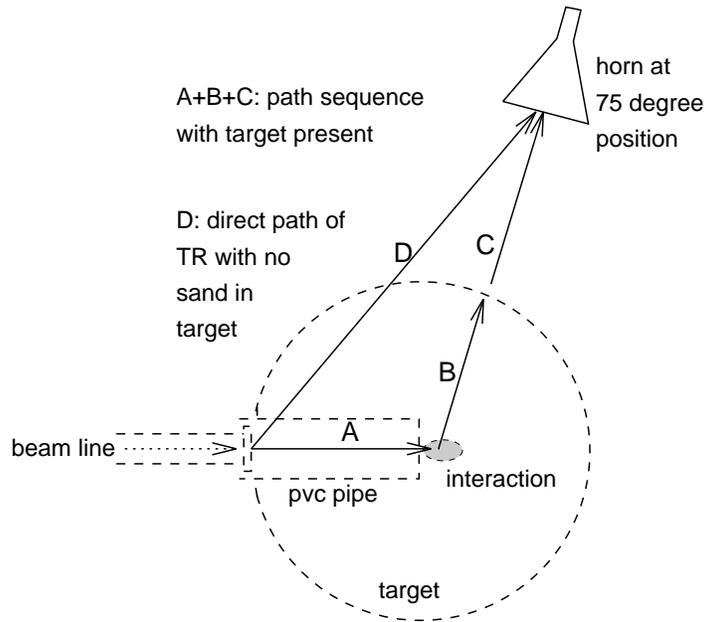}
\vspace{0.3cm}
\caption{An example of the time delay
geometry for two different cases: emission from
the end of the beampipe directly to the horn, when no sand is
present in the target; and the case of the beam
first interacting with the target and producing emission radiated
from its center to the horn, through the refractive target. }
\label{tdelay-fig}
\end{center}
\end{figure}

For example, in the $75^{\circ}$ profile, there
is a 2 ns delay of the full-target pulse relative to the empty-target
pulse. From Fig.~\ref{tdelay-fig}, the delay should correspond 
to the difference between the direct ray path from the beampipe
end to the horn (for the empty target case), and the composite
path in the full-target case, including: the beam path from beampipe
end to the target center, and then from the target center (where
the RF emission is formed) to
the horn. Thus the delay should be
\begin{equation}
\tau = c^{-1} ( A/\beta + nB + C - D )
\end{equation}
where $ A,B, C,D$ are the distances from
the end of the beampipe to the target center, the target center to the
target edge, the target edge to the horn face, and the beampipe end to
the horn face, respectively.
Since $A \simeq 70$~cm, 
$nB\simeq 1.5\times 40$~cm, 
$C=55$ cm, and $D=130$~cm, the expected delay is 1.75~ns,
consistent with what is observed, 
within the errors in our knowledge of the target position.
Note that the the polyethylene tub and PVC pipe
in the case where he target is empty have a negligible effect on
the pulse propagation since they have modest
microwave dielectric constants, and thicknesses much less than a
wavelength.

Based on the timing analysis presented here, 
we thus identify the time window between 12 and 15 ns as the
relevant portion of the plot in which to evaluate the possible
presence of a radiation component emanating from the
region near the target center. We will focus our attention
on the pulse energy in this time window in the sections that follow.

\subsection{Target-full angular power distributions}

With the target filled, we now expect production of RF emission
from the beam dump region. In addition, there is still
TR emitted from the beampipe end which
now passes through the target, where it is significantly
refracted by the cylindrical geometry and the refractive material
present. Thus although the target is designed to allow for radial
propagation of emission from near its center with minimal effects from
refraction; the refraction of emission not originating at the
target center will be significant and must be explicitly accounted for.
\begin{figure} 
\begin{center}
%\centerline{\psfig{file=tubtrace.eps,height=4.8in,width=4.8in}}
%\vspace{10pt}
\leavevmode
\epsfxsize=4.5in
\epsfbox{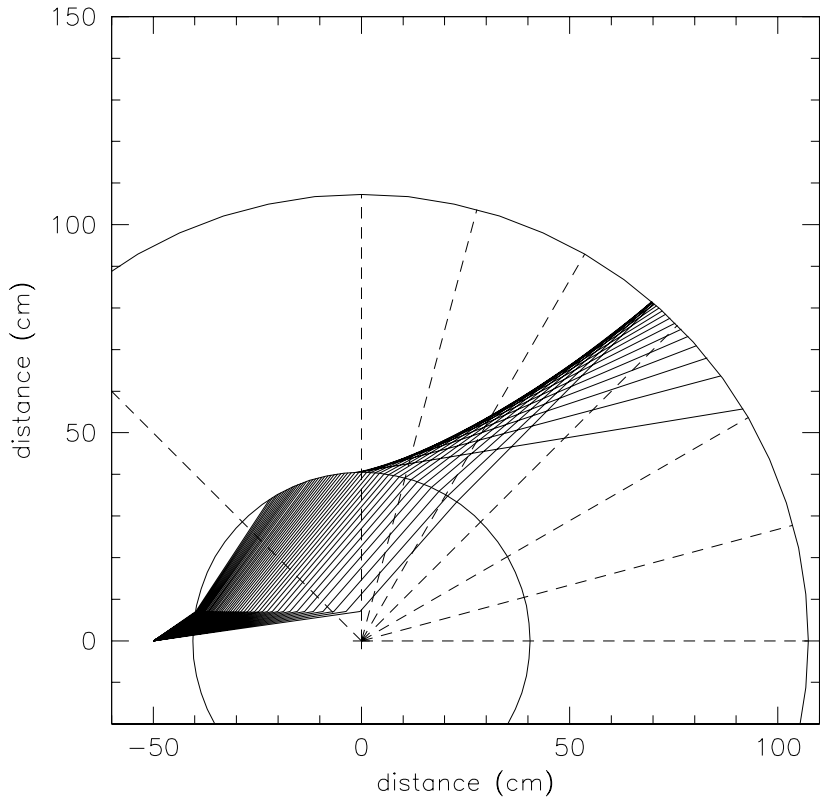}
\caption{Partial ray trace of the TR from the aluminum foil through
the sand target.}
\label{tubtrace-fig}
\end{center}
\end{figure}
Fig.~\ref{tubtrace-fig} shows the behavior of some of the rays
associated with the beampipe end TR, calculated using standard
geometrical ray-tracing equations. As the radiation enters the target
along the PVC input pipe wall, it is initially refracted 
away from the beam axis. Upon arriving
at the target boundary, however, the large change 
in the index of refraction from sand to air
tends to produce total internal reflection for rays at large angles, and
those that do escape tend to be highly forward-beamed. The net result
is that none of the TR from the beampipe end can propagate to angles
greater than $\sim 50^{\circ}$ with respect to the target center.
In fact the transmission coefficient at angles just above the
angle of total internal reflection are also quite small, and combining
this effect with the forward-peaked nature of the TR from the beampipe,
we find that there is virtually no contribution from the beampipe TR
beyond the $45^{\circ}$ position.
 
\begin{figure} 
\begin{center}
%\centerline{\psfig{file=new_flpwr.eps,height=4.8in,width=4.2in}}
%\vspace{10pt}
\leavevmode
\epsfxsize=4.5in
\epsfbox{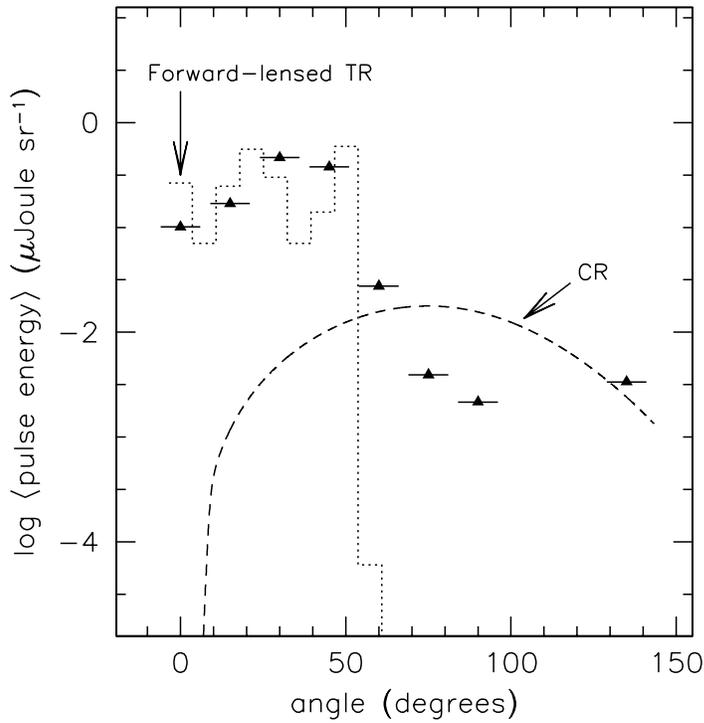}
\caption{The measured pulse energy (normalized by solid angle) 
at each angular position for the full
target is plotted along with the theoretically expected Cherenkov angular
distribution (dashed line);
Also plotted
is the expected distribution of lensed forward TR (dotted line), based on the
ray tracing described in the text.}
\label{flpwr-fig}
\end{center}
\end{figure}

Fig.~\ref{flpwr-fig} shows the measured pulse energy
(normalized to unit solid angle) at each of the angular
positions around the target-full case. 
We have also plotted (dashed curve) the expected CR angular
spectrum from the target center, using the measured
index of refraction of the sand and the predicted length of the
electron cascade near target center, taken here to be 4 cm.
This curve was
scaled down by a factor of 30 to match the
lower TR curve in figure~\ref{mtpwr-fig}. We have also
plotted an estimate of the angular intensity of
the forward-lensed TR (dotted lines). This curve has considerably more
uncertainty since it depends on both the accuracy of our ray-tracing, 
and on theoretical estimates for which the assumed conditions
(infinite media and tracklengths) do not apply. 

At angles smaller than $\sim 50^{\circ}$, the 
TR from the beampipe is focussed forward by the target and adds to
the CR from the target center. For $50^{\circ} < \theta < \sim 90^{\circ}$,
we expect little or no contribution of TR from the beampipe,
and we assume forward TR from
the quartz window is suppressed since the beam stops well within the
formation zone. Thus we expect CR to be the main contribution here.
At $\theta \gg 90^{\circ}$, both CR and backward TR from the quartz window
contribute, though again the TR undergoes some focusing which tends
to backward--beam it. We have not attempted to assess the 
backward TR in any quantitative way, but note that it
is likely to contribute to the increased pulse energy seen at
$135^{\circ}$ compared to 90$^{\circ}$.

%It is evident from this plot that, even allowing for the factor
%of $\sim 30$ discrepancy seen in the TR data, the CR power
%at the angles where it is expected to be dominant 
%appears to be $\leq 10$\% of the expectations based on the
%theoretical CR curve scaled down by the same deficit factor observed
%in the pure-TR runs.
It is evident from
the plot that we have not yet seen a clear signature of CR at expected levels.
In the remainder of this section we present other evidence which,
although somewhat indirect, does support the presence of 
CR emission in our observations, though it does not account for
the observed power deficit.

\subsection{Coherence at $60^{\circ}$}

\begin{figure} 
\begin{center}
%\centerline{\psfig{file=coh60.eps,height=4.6in,width=3.8in}}
%\vspace{10pt}
\leavevmode
\epsfxsize=4.5in
\epsfbox{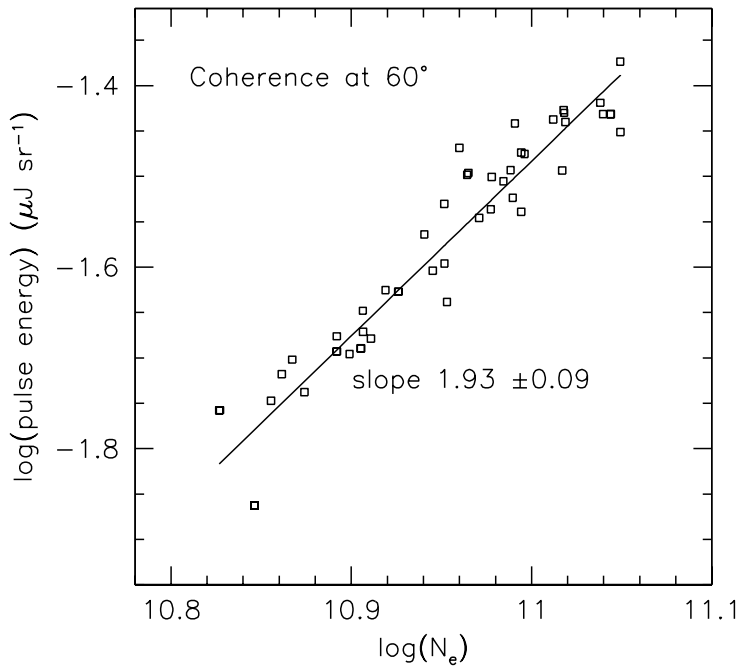}
\caption{Detected power plotted as a function of electron number in
the beam pulse, for two target-full runs at a horn angle of $60^{\circ}$.
A fit to the slope of the ensemble of points is also shown. Perfect
coherence would give a slope of 2.0, complete incoherence a slope of 1.0.}
\label{sndcoh}
\end{center}
\end{figure}

We are able to assess the coherence properties of the emission
observed at $60^{\circ}$ from the target center by investigating the
variation of observed power with the intrinsic variation of electron
bunch charge within a run.
Fig.~\ref{sndcoh} shows
a plot of the logarithm of the
integrated power at 60$^{\circ}$ within this window as a function of 
the logarithm estimated electron number in the beam pulse. The fitted slope
of this distribution is $1.93\pm 0.09$ which is consistent with
the emission being primarily coherent. Partial loss of coherence might be
expected in this case since the beam must propagate through 50 cm of air
path before it enters the target.  Note that this plot does not cover
as large a range of currents as Figure~\ref{trcoh}.

The observed level of coherence for these pulse charges is consistent
within errors with the overall coherence of the TR measurements
presented above, since these measurements were made at beam currents
which were at the high end of our range, where we had previously
seen a tendency for the coherence to roll off. This observed
coherence is of course expected from both TR and CR emission.

%\subsection{Spectral content as a function of angle}
%(To be added)
%

\subsection{Polarization dependence}

In the same way that we expect both CR and TR to be coherent
emission processes, they share similar polarization properties 
as well. In the limit of perfect coherence, we expect
both forms of emission to be completely linearly
polarized in the plane containing the beam velocity vector and
the direction of observation. For our observation geometry this
angle corresponds to $0^{\circ}$, or horizontal polarization.

A possible signature of a distinct emission component
could be indicated by a significant change
in the polarization state of the observed radiation
over a  particular range of angles. This
is due to the fact that at certain angular regions of our
measurements, the received radiation is a superposition of
TR refracted from the beampipe end (from different
ray paths), and possible CR from the target center. Such 
superposition of components of different phases
can be expected to produce noisy polarization 
measurements with a poorly defined plane of polarization.
In contrast, observations over an angular region where one
radiation component predominates should have less noise and
a more clearly defined plane of polarization. This is in fact
what we have observed.

Linear polarization measurements were made for four of the eight
angles around the full target. These measurements consisted of
recording the pulse profile at three different horn rotations around
the axis normal to its receiving aperture. The microwave
horn we employed was designed to have a principal E-plane, and had 
a cross-polarization rejection of $\sim 20$ dB for linear
polarization. Thus we can  estimate three of the four Stokes 
parameters $I,Q,U,V$ by making three intensity measurements at
$0^{\circ},~45^{\circ},~90^{\circ}$. Under these conditions, we have:
\begin{equation}
V_0^2 ~=~ {1 \over 2}(I+Q),~V_{45}^2 ~=~ {1 \over 2}(I+U),~V_{90}^2 ~=~ {1 \over 2}(I-Q)
\end{equation}
where $V_0, V_{45}, V_{90}$ are the voltage measurements at the specified 
angles. 
%Thus
%\begin{equation}
%I ~=~ V_0^2 + V_{90}^2,~Q ~=~ V_0^2 - V_{90}^2,~ 
%U~=~V_{45}^2-{1\over 2}(V_0^2 - V_{90}^2)
%\end{equation}
and these results can then be combined in the usual way to give the
fractional polarization $P$, and the polarization angle $\phi$:
\begin{equation}
P ~=~ {\sqrt{Q^2 + U^2} \over I} ; \ \tan \phi ~=~ {U \over Q}~. 
\end{equation}

\begin{figure}
\begin{center}
%\centerline{\psfig{file=Stokesall.eps,height=5.2in,width=4.6in}}
%\vspace{10pt}
\leavevmode
\epsfxsize=5in
\epsfbox{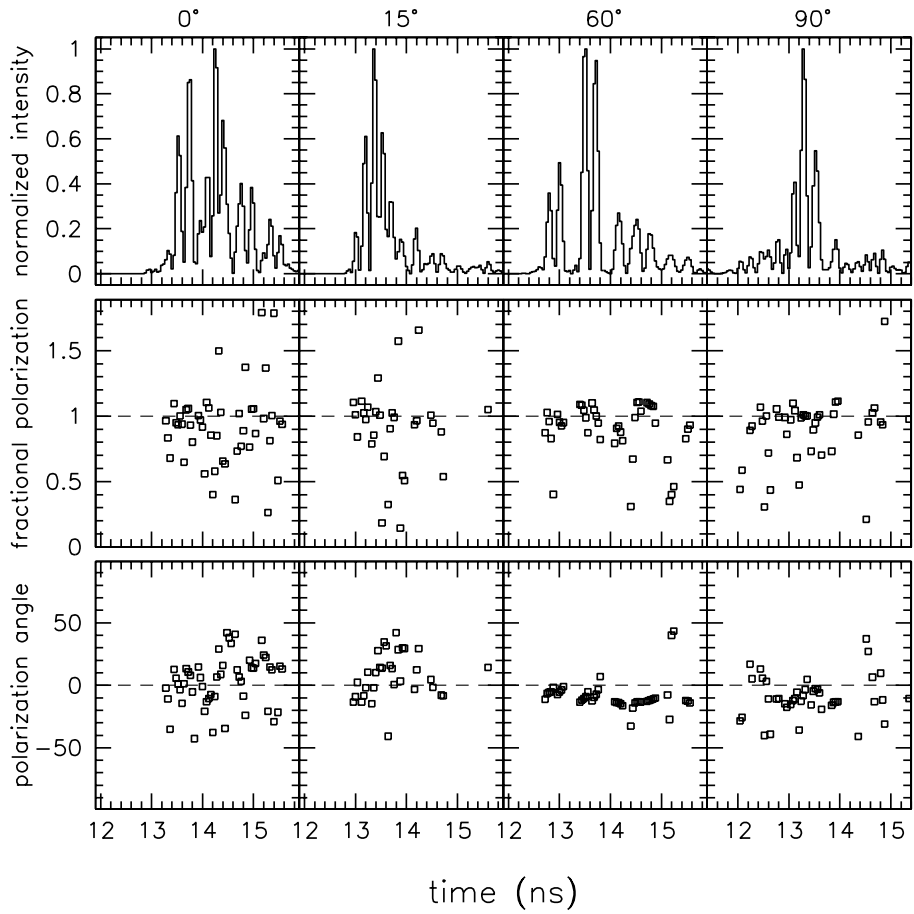}
\caption{The normalized power 
profile in the primary pulse is plotted along the top
row, followed by the fractional polarization and the polarization
angle, for four positions measured around the target. The fractional
polarization and corresponding angle are only plotted where the
power is greater than 0.05 of the maximum. The resulting 
scatter is thus intrinsic, not
statistical. The plane of polarization is expected to be $\sim 0^{\circ}$
for direct CR or TR. }
\label{pol-fig}
\end{center}
\end{figure}

The results for the four positions measured are shown in Fig.~\ref{pol-fig}.
The different angle positions around the target are plotted
columnwise, and the intensity profile for each case is displayed along the
top row, followed by the fractional polarization and the polarization 
angle along the bottom of the plot. For the fractional polarization and
angle, only points with intensities greater than 0.05 of the maximum are
plotted; thus each point has relatively high SNR, and the
observed scatter is intrinsic rather than statistical. For TR and CR
that arise from a single location and are unscattered, we expect
the fractional polarization to be $\sim 1$ and the plane of
polarization to be horizontal ($0^{\circ}$). 

We find that the characteristics of the polarization at angles greater than 
$50^{\circ}$ are notably different than those of the
forward angles where we expect significant superposition of different
forward scattered and reflected components of strong TR. At the
$0^{\circ}$ position with respect to the target, a somewhat larger
scatter in the angle of the plane of polarization is to be expected,
due to the effects of being close to the polar axis.
In contrast, the 
$15^{\circ}$ position with respect to the target should give a more
clearly defined plane of polarization but also shows a large scatter,
consistent with mixing of multiple components in the TR.

\begin{figure}
\begin{center}
\leavevmode
\epsfxsize=5in
\epsfbox{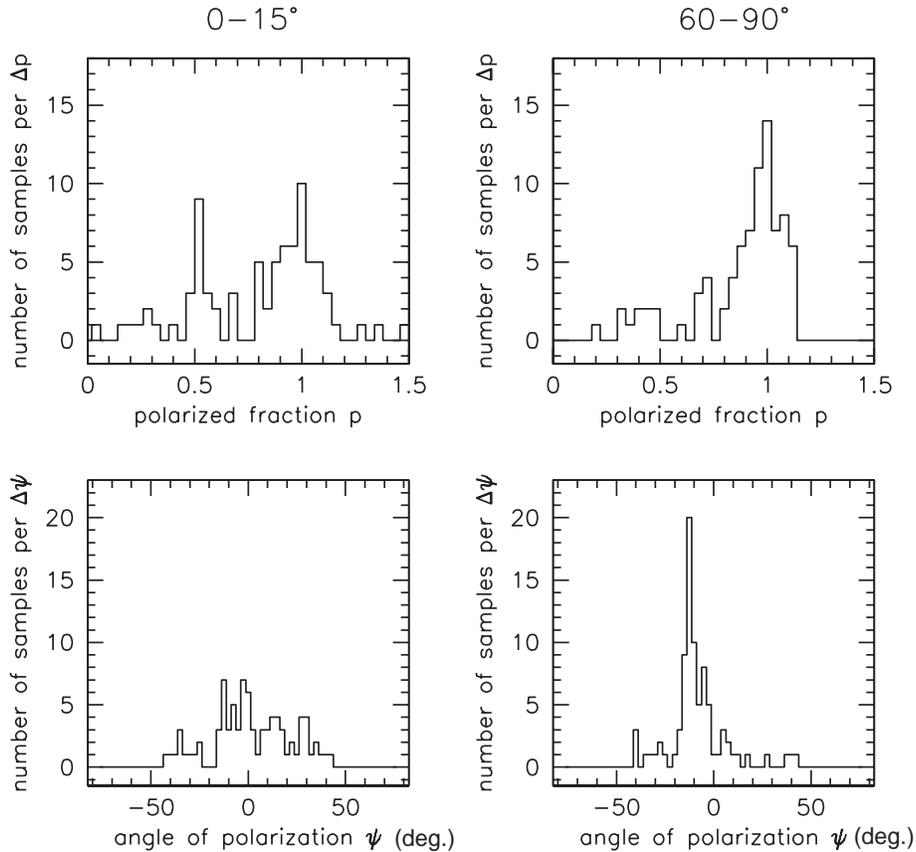}
\caption{Histograms of the sampled fractional polarization and
angle of the plane of polarization, for two different
angular regions, $0-15^{\circ}$ and $60-90^{\circ}$, showing the
distinct behavior for the larger angles, where Cherenkov
emission is expected to contribute.}
\label{polhist-fig}
\end{center}
\end{figure}

At $60^{\circ}$ we see a markedly different behavior in the plane
of polarization, with much less scatter and values that are close
to $0^{\circ}$. This behavior is also present but somewhat weaker at
$90^{\circ}$. This behavior is displayed more 
clearly in Fig.~\ref{polhist-fig},
where we have now combined the data at the low angles for
comparison with the data at the higher angles. It is evident that
both the fractional polarization and the plane of polarization are
more clearly defined at the larger angles.\footnote{The apparent
$\sim 12^{\circ}$ offset in the plane of polarization from 
zero is likely due to systematics in our experiment; we had
no independent calibration of the polarization.}
We interpret this as evidence that we are seeing a
single coherent radiation component that displays characteristics
consistent with Cherenkov radiation from the target center. We
cannot rule out a residual contribution of TR from the
beampipe end here, but as we have noted above, at these
larger angles we do not expect to see significant TR.

We have also considered the possibility that we are seeing
a wide-angle forward TR from the air-quartz-sand interfaces, and little or
no CR.  Although we cannot rule this out we note that
the bulk of the electrons stop well within the 
formation zone (estimated to be 20-30cm
at 60$^\circ$) and we expect that the TR is thus suppressed.  In 
addition, the presence of the sand surrounding the beam (at a radius 
of $\sim 8$ cm --- less than 1 wavelength) as it 
traverses the target prior to penetrating the quartz window also
``smooths'' the dielectric transition and will further tend to 
suppress the TR.
We note also that theoretical predictions for the 
power and angular spectrum of TR
for this interface are inconsistent with energy considerations,
as we show in appendix A; so we are unable to make
reliable estimates of the power that might be present in this case.

\section{Discussion}

In the previous section we have presented results which show
a clear signature of coherent transition radiation at GHz radio
frequencies from electron bunches. Although this result is not
surprising given similar results at far infrared and submillimeter
wavelengths, it is to our knowledge the first such demonstration at
these longer wavelengths.

In addition, we show evidence consistent with a Cherenkov
component in the emission from our target. This evidence relies
primarily on the design of the target, which produces an
angular window where the TR emission is suppressed, and is
strengthened by the presence of distinct, linearly polarized
component in the emission near the Cherenkov angle, which contrasts
sharply with the TR seen at other angles.
However, we do not yet observe either
Cherenkov or transition radiation that is consistent
with the expected power in our experiment. 

One possibility is that complete coherence does not fully obtain in
either our TR or CR production, due to possible saturation 
or self-quenching effects in the electron bunch. There is in fact some
evidence that a partial loss of coherence is present
in the TR runs (see figure~\ref{trcoh}).
This effect appears to set in only at the highest beam charge values.
As discussed in the following subsection such a roll-off for
the highest energy showers must happen at some point.

\subsection{Implications for cascade radio emission}

If we take the value of $N_e$ as a measure of the
charge excess in a cascade, we can associate our measurements with
an equivalent cascade energy $E_e \simeq N_e/0.2$ GeV, where we
have assumed here that the charge excess is 20\% of the total 
charge near shower maximum. For the bunch charges in our
experiment, $E_e \simeq 5\times 10^{20-21}$ eV, corresponding to 
energies that equal and exceed the highest energy cosmic ray cascades
detected to date.

Zas, Halzen, and Stanev~\cite{Zas92} have shown that high energy
cascades in solid ice, under the assumption of full coherence,
produce a total radio CR pulse energy $W$ which can be written as
\begin{equation}
\label{cascade-eq}
W ~\simeq~ 5 \times 10^{-15}~
\left ( {E_e \over 1 ~{\rm TeV} } \right )^2
\left ( {\nu_{max} \over 1 ~{\rm GHz} } \right )^2
~~{\rm \mu J}~
\end{equation}
where $\nu_{max}$ is the highest frequency observed (in practice
loss of coherence sets in for $\nu_{max} \geq 2$ GHz).
This quadratic dependence of radio power implies a saturation
energy: that is, we can equate the total energy in equation \ref{cascade-eq}
with the total energy in the shower to derive an upper limit 
$E_{max}$ for the maximum 
energy at which this relation can be valid. Thus we find
\begin{equation}
E_{max} ~=~ 3 \times 10^{24}~{\rm eV}~.
\end{equation}
which is only
a factor of $600$ above our highest equivalent cascade energy.
However, $E_{max}$ represents the value at which {\em all} of
the shower energy is lost to the RF pulse; in fact a more conservative
approach would require some type of equipartition of the shower
energy among other energy-loss mechanisms such as ionization,
and would require that the radiation reaction of the shower
conserve total momentum.
Thus we predict that equation \ref{cascade-eq} must begin to lose
validity at some critical energy $E_{crit} = \epsilon E_{max}$, where
we expect that $\epsilon \leq 0.1$. Our data in fact suggest 
$\epsilon \sim 0.001$ if the loss of coherence we observe at the
highest bunch charges is due to some related effect.

Similar arguments can be applied directly to the theoretical power
expected in our experiment from both CR and TR. If we integrate
over just the forward angles for our experimental conditions, the
total expected power in our passband is $\simeq 60$ $\mu$J. The
total energy per electron bunch is $N_e E_e \simeq 2.2 \times 10^5$
$\mu$J for 15 nC and 15.2 MeV electrons. Thus the theory predicts that about
1 part in 3600 of the bunch energy to be radiated in our band
for fully coherent emission. However, this implies that only a factor
of 60 increase in the bunch charge would lead to a complete loss
of energy of the bunch to coherent RF emission. Thus it appears
that our experimental condition may be more severely affected by
possible loss of coherence than equation \ref{cascade-eq} implies.

\subsection{Lower limits on cascade power at $E_0 \geq 10^{20}$ eV }

In spite of the difficulty in finding a satisfactory resolution to
the apparent discrepancies in observed vs. theoretical power,
we can still use our observed values to derive lower limits on the
expected power from both TR and CR from cascades in materials 
similar to what we have observed. For TR, refractory materials
such as sand, lunar regolith, or ice produce nearly the same
spectrum as the aluminum window used on our beampipe end. Thus
our results are directly applicable to cascades that encounter
interfaces in these materials, for example, a cascade
that emerges from the material into either air or vacuum is
closely analogous to what we have observed for TR.
In the CR case, our results are more uncertain but may still 
provide a preliminary experimental
lower limit in estimating cascade emission.

\subsubsection{Transition radiation}

The highest values in our TR measurements of the pulse
energy were $\sim 6$ $\mu$J sr$^{-1}$. Converting these
to the flux density units commonly used in radio astronomical 
measurements, we estimate that the TR flux density produced at earth
from a cascade of $E_0 = 5 \times 10^{21}$ eV
emerging from the surface of the lunar regolith is
\begin{equation}
S (\theta \simeq 10^{\circ}) ~\geq~ 5000  
\left ( {E_0 \over 5 \times 10^{21} ~{\rm eV}} \right )^{1.9} ~{\rm Jy}
\end{equation}
where $1$ Jy $\equiv 10^{-26}$ W m$^{-2}$ Hz$^{-1}$. Since the TR
angular distribution appears to be confirmed by our data for
these conditions, we can also estimate that the maximum
flux density for this case, at an angle of $\sim 1.5^{\circ}$
from the cascade axis, is about a factor of 20 higher than at $\sim 10^{\circ}$:
\begin{equation}
S_{max} (\theta \simeq 1.5^{\circ}) ~\geq~ 10^5 
\left ( {E_0 \over 5 \times 10^{21} ~{\rm eV}} \right )^{1.9}~{\rm Jy}~.
\end{equation}
The large antennas used in lunar pulse searches have 
demonstrated the ability to achieve noise levels of 
order 400~Jy~\cite{Gor99} or less.
Thus it appears that the energy threshold for detection of events
by TR alone is $5 \times 10^{20}$ eV. Here we have not
considered the indications of higher measured pulse energy 
given by our dipole measurements. If these results are correct,
the energy threshold may be an order of magnitude lower.

These results support the claim~\cite{Mark86} that TR detection of cascades
is feasible for the case that a cascade emerges through some surface
into a medium transparent to radio emission, and this analysis
also applies to detection of cascades on earth. For example, our measurements
could be applied to the case of a shower that was initiated
in air and then entered ice near a subsurface radio array, such as
has been proposed and prototyped in Antarctica~\cite{fri96,Bes99,Seck99}.
In this case, we can write the threshold energy more generally as
\begin{equation}
E_{thr}^{TR} ~\leq~ 5 \times 10^{20}
\left ( {\Delta \nu \over 900~{\rm MHz}} \right )^{-1/2}
\left ( {A_{eff} \over 2500~{\rm m^2}} \right )^{-1/2}
\left ( {  R  \over 3.8\times 10^8~{\rm m}} \right )
\left ( {T_{sys} \over 220~{\rm K}} \right )^{1/2} ~{\rm eV}.
\end{equation}
where $A_{eff}$ is the total effective collecting area of the antenna
or antenna array
(including aperture efficiency), $R$ is the distance to the
cascade, and $T_{sys}$ the system temperature. Here we have assumed that 
full coherence obtains at lower cascade energies; thus the
exponent for the energy term above was assumed to be 2.0 rather
than 1.9 as we have used at the highest energies.

If we apply this equation to the case of an antenna
array in ice comprising 10 m$^2$ effective aperture,
operating at $T_{sys}$ of 1000 K with a 900 MHz effective bandwidth, 
the system will have a TR detection 
capability for cascades with $E_0 \simeq 2 \times 10^{16}$ eV 
entering the ice from above, out to a distance of 250 m. 
However, we caution that formation zone effects may be important
depending on how rapidly the near-surface refractivity of the
ice changes with depth.

\subsubsection{Cherenkov radiation}

Although we attempted to optimize our experiment in favor or
CR detection rather than TR, we cannot yet claim to have yet made 
measurements of it adequate to allow us to make definite statements
about the application of these data to the detection of radio CR from
high energy cascades. One thing is much more clear to us
after having performed this initial experiment than was evident
in the literature: the processes that produce these two forms of
radiation are physically very closely aligned.

The problem of separating the two forms of emission is particularly
acute when the radiation can be formed in nearly the
same physical location, such as at the quartz window entering the
target in our case. In fact, the theoretical formulations for
TR at this point do not themselves clearly separate the two processes;
as we noted in an earlier section, ``Cherenkov-like'' components
appear even in the TR equations. Thus, we suggest that a more
complete theoretical basis for TR formation under realistic
conditions with finite track lengths and boundaries be developed.

An important aspect of the operating regime of our experiment
is that the detected Cherenkov
power should scale quadratically with both
the total track length as well as the number of particles.
For solid materials, the mean electron energy near shower maximum
for a typical high energy cascade is closer to
$\sim 100 $ MeV rather than the 15 MeV used in our experiment.
Thus the total track length is much longer and the Cherenkov 
production is more directive and therefore more intense near its
peak angle. Any scaling from experiments similar to
ours must account for this effect.

If we do attribute the observed power 
of 0.014 $\mu$J sr$^{-1}$ at our $60^{\circ}$ angle
primarily to Cherenkov radiation, we can make a tentative estimate of
the possible energy threshold for CR detection.
A cascade of energy $\sim 10^{20}$ eV in ice or the lunar regolith
will have a length of order 10 m\cite{Zas92} near shower maximum. 
For this length, the implied angular enhancement
in the intensity is of order $10^4$, giving $\sim 140~\mu$J sr$^{-1}$,
comparable to the observed TR intensity
at an angle of $\sim 1.5^{\circ}$.
The implied cascade energy threshold, using the same arguments
as above, is therefore comparable to that implied by our TR
results. For the lunar regolith, $E_{thr}~\approx~ 5 \times 10^{20}$ eV,
similar to values estimated from complete electromagnetic
simulations of ultra-high energy cascades~\cite{Zas92,Alv96,Alv97}.
If our assumptions are correct, this value is conservative,
and we expect further experiments to yield a potentially much
lower cascade energy threshold for coherent Cherenkov detection.

\subsection{Implications of RF lensing effects}

In concluding this section we note that the lensing effects
on the forward-beamed TR emission that we have observed
in our target-full configuration have potentially important
implications for experiments to detect cascade emission from
the lunar regolith or other materials 
observed through a refractive interface. Under these conditions,
our results suggest that
total internal reflection of the cascade radio emission
can strongly suppress it at certain angles and under certain
geometrical configurations.

This effect is potentially helpful in distinguishing
particles such as hadronic cosmic rays or photons
(which interact immediately upon entering the 
refractive material) from neutrinos
(which can interact near the surface of the material after
traversing it over large distances).
In the latter case, since
the critical angle for total internal reflection
is the complement of the Cherenkov angle at a
vacuum interface,
the CR can exit the
surface into the vacuum.  However, in the former case, the CR from the
cascade will suffer total-internal reflectance and
will not emerge from the surface.
Clearly, surface irregularities at scales approaching
an RF wavelength will modify this conclusion at some
level, but to first order this effect will tend to suppress
the detection of cosmic ray events relative to upcoming neutrino
cascades in lunar observations.

\section{Future work}

Our ability to separate TR from CR in this analysis depended on the
geometry and precise timing.
TR from the vacuum window was lensed into most of
our observing angles.  In a new experiment, with the window further
forward, much of this effect will be avoided. In addition, taking
more position angles with full polarization measurements will further assist
us in separating prompt from lensed power.

Our results suffer in part from the difficulty of achieving
accurate power calibration. In future work, we will significantly
improve our power calibration capabilities.
We would also plan to more carefully calibrate the beam current measurements
with a Faraday cup to determine if the apparent loss of coherence
that we see at high beam currents is real or due to nonlinearities
in our measurement. 

Future experiments along these lines should also be done using
pulsed high energy photons, whose radiation length in sand
(30--40~g~cm$^{-2}$) will allow the shower to develop {\em within}
the target and will thus avoid the generation of transition
radiation at the beampipe end and entrance to the target.

On the theoretical side, analytic or parametric formulas for dealing with 
disturbances of the charge inside the TR formation zone would be helpful.
We would also like to see a theoretical understanding of the critical
energy beyond which the coherence fails to scale at $N_e^2$ which must
be important near the energies and currents we are using.

\section*{Acknowledgments}

We thank George Resch and Michael Klein for their 
generous support of this work, Michael Spencer for the
excellent S-band dipoles he produced, and John Ralston, 
Jaime Alvarez-Mu\~niz, and Enrique Zas for their helpful
comments on the manuscript.   We thank Jamie Rosenzweig for valuable
advice.
The Argonne Wakefield Accelerator is supported under U.~S. Department
of Energy, Division of High Energy Physics contract 
W-31-109-ENG-38.
This research has been performed in part at the Jet Propulsion
Laboratory, California Institute of Technology, under contract with
the National Aeronautics and Space Administration.
This work was also supported in part by the A.~P.~Sloan Foundation and
by the UCLA Division of Physical Sciences.

\appendix
\section{Predictions for radio-frequency coherent TR}
\label{examples}

To illustrate some of the complexity in evaluating the theoretical
predictions for TR in the radio frequency regime, we have calculated
the forward angular spectra for TR in a number of configurations relevant
to our experiment, using equation~\ref{tr-eq}. We have
chosen a frequency range of 1.7--2.6 GHz, corresponding to
the range used in our experiment. We have considered emission from
a single bunch of 15 MeV electrons with a total charge of 15~nC.
For our purposes here we have scaled the results
by a factor of $N_e^2$, corresponding to perfect coherence.

Figure~14 shows the results of this analysis for
several combinations of upstream and downstream materials,
noted in the panel text. In each case we have plotted the angular
spectrum predicted by equation~\ref{tr-eq} with a solid line.
Also plotted (dashed curve) is the integral over the
angles shown  of the total energy of the pulse. The dotted curve shows
the total available kinetic energy of the electron bunch.

In Fig.~14a we plot predictions for the case
most commonly encountered in the experimental literature for
submillimeter and far-IR measurements of TR from electron
bunches: radiation from the exit of the bunch through
a beampipe metal window into air. 
In this case TR is forward-peaked at an angle of
$\sim 1/\gamma$, and the total energy radiated in coherent
TR in this case is $\sim 2\times 10^{-4}$ of the available
kinetic energy. As we have seen above, we find that our data
confirms the angular distribution shown here, though not the total
pulse energy.

In Fig.14b, we plot the prediction for
the case of the electron bunches exiting a metal window directly
into a dielectric material with $n=1.5$ corresponding to silica
sand as we have used in our experiment. Here the strong peak
at the nominal Cherenkov angle is evident, and the forward
peak seen in the previous pane is no longer present. Also
important is the fact that the total energy in the
coherent emission now approaches 10\% of the available kinetic
energy. This prediction thus appears to be physically
improbable.

In Figs.~14c and d we see even more complex behavior in
the cases of a sand-to-air and air-to-sand transitions. Clearly
in both cases the predicted total energy exceeds the
available kinetic energy and thus the theory appears to
be inadequate in treating this combination of parameters.

This example highlights two issues that we have faced in this experiment: 
First, what is the actual TR angular
distribution in cases that differ from the simple
metal-to-air interface commonly treated by the theory? 
Second, how do we resolve the issue of what fraction of the
total energy in a particle bunch can reasonably be emitted
into coherent radiation?

\begin{figure}
\begin{center}
\leavevmode
\epsfxsize=5.5in
\epsfbox{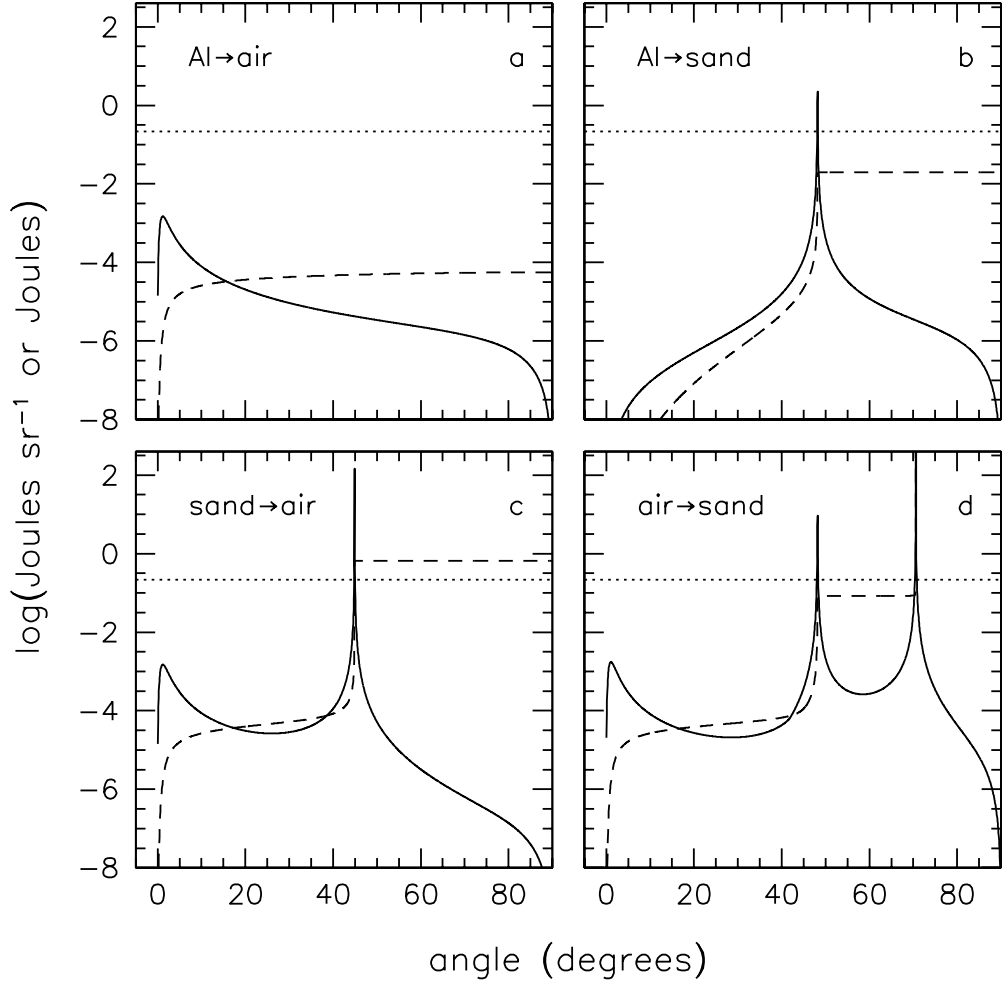}
\label{tr-fig}
\caption{
Predictions from equation~\ref{tr-eq} for emission of
coherent radio frequency TR using various dielectric combinations.
The wavelength range used in 11-18 cm, and the electron bunches
have an energy of 15 MeV with a bunch charge of 15 nC. In each
case the solid curve gives the angular spectrum, the dashed curve
the integral of that spectrum, and the dotted curve the total
available kinetic energy of the bunch. (a) The metal-to-air interface. 
(b) The aluminum-to-sand interface
predicts a Cherenkov-like distribution which appears to emit
too large a fraction ($\sim 10$\%) of the total energy into TR.
(c\& d). The sand-air and air-sand interfaces predict a complex behavior
which is physically unrealistic since the total energy exceeds that
of the available kinetic energy.}
\end{center}
\end{figure}

\end{document}